# Variation in research collaboration patterns across academic ranks[1]

Giovanni Abramo
*Laboratory for Studies of Research and Technology Transfer
at the Institute for System Analysis and Computer Science (IASI-CNR)
National Research Council of Italy*

ADDRESS: Istituto di Analisi dei Sistemi e Informatica, Consiglio Nazionale delle Ricerche, Viale Manzoni 30, 00185 Roma – ITALY
tel. +39 06 7716417, fax +39 06 7716461, giovanni.abramo@uniroma2.it

Ciriaco Andrea D'Angelo
*University of Rome "Tor Vergata" – Italy and
Laboratory for Studies of Research and Technology Transfer (IASI-CNR)*

ADDRESS: Dipartimento di Ingegneria dell'Impresa, Università degli Studi di Roma "Tor Vergata", Via del Politecnico 1, 00133 Roma – ITALY
tel. and fax +39 06 72597362, dangelo@dii.uniroma2.it

Gianluca Murgia
*University of Siena – Italy*
ADDRESS: Dipartimento di Ingegneria dell'Informazione e Science Matematiche, Università degli Studi di Siena, Via Roma 56, 53100 Siena – ITALY
tel. and fax +39 0577 1916386, murgia@dii.unisi.it

# Variation in research collaboration patterns across academic ranks

**Abstract**

The ability to activate and manage effective collaborations is becoming an increasingly important criteria in policies on academic career advancement. The rise of such policies leads to development of indicators that permit measurement of the propensity to collaborate for academics of different ranks, and to examine the role of several variables in collaboration, first among these being the researchers' disciplines. In this work we apply an innovative bibliometric approach based on individual propensity for collaboration to measure the differences in propensity across academic ranks, by discipline and for choice of collaboration forms - intramural, extramural domestic and international. The analysis is based on the scientific production of Italian academics for the period 2006 to 2010, totaling over 200,000 publications indexed in Web of Science. It shows that assistant professors register a propensity for intramural collaboration that is clearly greater than for professors of higher ranks. Vice versa, the higher ranks, but not quite so clearly, register greater propensity to collaborate at the international level.

# 1. Introduction

One of the critical competencies for academics is the capacity to develop effective collaborations (Bozeman and Corley, 2004). Through collaboration, individual academics can build up their research activity and gain the benefit of the results (Jeong et al., 2011; Abramo et al., 2009). Moreover they will also assist the growth of their home institution, since successful collaboration assists in ensuring flows of research funding (Traore and Landry, 1997), strengthens the university's scientific reputation and contributes to attracting talented PhD students and post-docs (Casey et al., 2001). For these reasons, the capacity to construct effective collaborations is often identified as one of the base criteria for career advancement (Jeong et al., 2011).

While propensity for collaboration favors career progression, the reverse can also be true (van Rijnsoever et al., 2008). In fact, the capacity to construct effective collaborations is related to the management of tasks, such as the creation and management of research groups, management of laboratories and complex structures, and responsibilities in departmental and university administration (Zuckerman and Merton, 1972), that for university regulations and tradition are assigned to the academics of highest rank.

In the past, even if a scientist had only developed a network of collaborations at the domestic level this could be considered sufficient for promotion to full professor. However, it is now commonly expected that the network of collaborations should be well developed to at least the continental level (Ackers, 2004), in part because the large share of research funds is obtained through international competitions (Arthur et al., 2007).

It is no accident that top scientists show a greater propensity for international collaboration than others (Abramo et al., 2011a). The propensity to collaborate can also be influenced by other inter-related variables: i) age (Kyvic and Olsen, 2008), which is broadly related to rank due to the prerequisite of experience for higher positions (Knodt and Kotzian, 2009; Lee and Bozeman, 2005; Long et al., 1993); ii) gender (Abramo et al., 2013a), which appears particularly influential in incidence of international collaborations, where demands on mobility may penalize women of family-rearing age, which is also when female researchers tend to hold lower ranks; iii) the researcher's scientific discipline (Abramo et al., 2013b), where the link with rank is quite complex. For example, in Physics, given its nature as a "big science", an international collaboration network is considered essential even for access to the lower ranks (Ackers, 2005).

This work deals with the relationship between academic rank and intensity of collaboration in research. In particular, we wish to verify if there are differences in the propensity to collaborate for full, associate and assistant professors in terms of the various forms of collaboration (intramural, extramural domestic, international), and whether these differences vary across disciplines and fields. The field of observation is made of publications indexed by Web of Science (WoS). Although they are not the only form of output from scientific collaboration, co-authored publications provide a meaningful proxy. Assumption of this proxy also offers advantages of ease in measurement and freedom from biases related to the object of analysis or subject conducting the observations (Melin and Persson, 1996). Thus we adopt a bibliometric approach, employing a methodology in which the single scientist is the base unit of analysis. For the Italian higher education system, we are able to attribute each publication to the academic that produced it, with a very low error rate. Through the scientists' affiliation, we are then able to discriminate intramural from extramural collaborations and domestic extramural from international collaborations. To evaluate whether the differences in propensity to collaborate between the three ranks vary by discipline or field, we take advantage of a unique characteristic of the Italian university system, where each academic is classified as belonging to one and only one field (Scientific Disciplinary Sector, SDS). There are 370 such fields[2], in turn grouped into 14 disciplines (University Disciplinary Areas, UDAs).

The empirical evidence from our study can help the policy maker in the design and evaluation of policies for management of collaboration that take into account the levels of experience and motivations of the different academic ranks (van Rijnsoever et al., 2008).

In Section 2 of this paper we discuss the relevant literature, and then describe the methodology applied and the field of observation in Section 3. Section 4 presents the results of the analyses, while the final section proposes several indications for policy and future directions for study.

## 2. Relationships between academic rank and research collaboration: literature review

In the literature, the relationship between scientific collaboration and academic rank has primarily been studied through comparing the situations of academics and young non-faculty researchers (Hinnant et al., 2012; Stvilia et al., 2011; Hagstrom, 1965), or those of tenured and untenured researchers (Fox and Faver, 1984; Bozeman and Gaughan, 2011). Still, some studies have indicated that variation in forms of collaboration can develop among the different ranks of tenured researchers, since the circumstances of full, associate assistant professorship give rise to specific and distinct motivations in this regard (Rivellini et al., 2006; Lee and Bozeman, 2005).

One of the primary motivations for individual scientists to collaborate is to increase their scientific production (Beaver, 2001; Liberman and Wolf, 1998). This motivation occurs for all academics regardless of their rank, although some full professors, particularly when near the close of their careers, could be less concerned about scientific productivity given the lesser impact on their personal prospects (Jeong et al., 2011; Kyvic and Olsen, 2008). In a related manner, especially in the social sciences, academics of lower ranks could be less inclined to collaborate (Vafeas, 2010) because of the perception that publication of co-authored articles might limit recognition of their capacities as individual scientists (Mcdowell and Melvin, 1983; Piette and Ross, 1992), thus hampering their career prospects (Fox and Faver, 1984; Siva et al., 1998). This motivation against collaboration is exacerbated by awareness of the "Matthew effect", which holds that the merit for a co-authored article will go primarily to the most famous of the scientists credited in the byline (Merton, 1968).

On the other hand, the link between scientific production and career advancement (Lissoni et al., 2011) can push academics of lesser rank towards collaboration, since the adoption of co-authorship increases the probability of success for a publication (Presser, 1980). In particular, collaboration with more expert colleagues helps the scientist to improve the quality of their works prior to submission for publication (Barnett et al., 1988; Beaver, 2001). Collaboration with better known colleagues within a discipline can also gain stronger "sponsorship" in proposals to editors and among referees (Baethge, 2008; Petty et al., 1999; Laband and Tollison, 2000), as well among colleagues who could then cite the publication (Hinnant et al., 2012).

---

[2] The complete list is accessible at http://attiministeriali.miur.it/UserFiles/115.htm, last accessed on Aug. 30, 2013.

It is increasingly common practice for higher ranking colleagues to receive "remuneration" in the form of inclusion in co-authorship of articles. "Gift authorship" occurs when their true contribution to research would not justify inclusion in the byline (Smith, 1994; Bayer and Smart, 1991). This phenomenon occurs primarily in the disciplines known as "big sciences" (Street et al., 2010), arising particularly from mentorship arrangements (Baethge, 2008; Kwok, 2005; Bayer and Smart, 1991) and in research that requires special equipment and samples (Ezsias, 1997; Price, 1963). Gift authorships tend to give rise to greater benefits for the higher ranking researchers (Drenth, 1998) and indeed senior academics may actually develop such arrangements by exercising coercive power over younger researchers (Kwok, 2005).

Apart from such extreme cases, the identification of authorship for higher ranking academics is often linked to the fact that they can more easily take on searches for funds (Street et al., 2010; van de Sande et al., 2005), manage projects and laboratories (Baruch and Hall, 2004; Bayer and Smart, 1991; Bordons et al., 2003) and attract greater numbers of fellows (Martin-Sempere et al., 2008; Luckhaupt et al., 2005). The greater responsibilities and the related greater resources for senior academics mean that they also tend to develop collaboration networks that are broader (Bozeman and Gaughan, 2011), more cosmopolitan (Bozeman and Corley, 2004; Zuckerman and Merton, 1972), consolidated and productive (Martin-Sempere et al., 2008). In fact, higher ranking researchers may increase their collaborations to favor the growth of their professional fellows and to ensure adequate management of their research projects (Fox and Faver, 1984; Rivellini et al., 2006). On the other hand, as also postulated under cumulative advantage theory (Merton, 1968), many collaborations that involved higher-ranked academics are due to the need for access to resources, such as laboratories, equipment, and administrative support personnel. Without the involvement of senior academics, those of lower rank could be completely excluded from such resources. Indeed it is natural that mentorship relationships generally involve full or associate professors in the mentor role and assistant professors in the role of "mentee" (Sands et al., 1991). In this manner, assistant professors can take advantage of the greater experience (Rivellini et al., 2006) and social capital (Etzkowitz et al., 2000) of the senior academics. Lower ranking academics are also pushed to collaborate, not only to overcome the gap in availability of resources but also to demonstrate their capacity to activate and manage collaborations, which are considered essential to career progress (Traore and Landry, 1997; Bayer and Smart, 1991).

Activation and management of collaborations implies costs that that can vary significantly by rank of the academics involved. This inequality may in some cases be due to the unequal division of duties in the collaboration, at greater expense to researchers with less power. However, more frequently the differences in tasks would be because of the different levels of experience of full, associate and assistant professors (Lee and Bozeman, 2005).

As we would expect, lower ranking academics seem primarily involved in collaborations at the intramural level, while collaborations that involve higher numbers of organizations see a strong presence of senior academics (Hinnant et al., 2012). Still, in absolute numbers, full professors register a higher number of intramural collaborations than do lower-ranking academics, thanks in part to having a greater number of fellows (Lee and Bozeman, 2005; van Rijnsoever et al., 2008). The greater number of extramural collaborations for full professors is in part favored by their potential for involvement in governance activities, which permits them to activate links with colleagues from other universities, particularly at the domestic level (van Rijnsoever et al., 2008).

In regards to collaboration at the international level, full professors register more events than associate and assistant professors (Frehill et al., 2010), particularly if they have access to substantial research funds and are lower in seniority (Melkers and Kiopa, 2010). The international collaborations of full and associate professors are activated particularly through conferences, while those of assistant professors are very often linked to participation in foreign PhD programs (Melkers and Kiopa, 2010). The lesser number of international collaborations for assistant professors is also due to the fact that these have less relevance for access to higher ranks (Arthur et al., 2007), with the exception of certain disciplines, such as in Physics (Ackers, 2004; Ackers, 2005).

The observations of differences in motivation and costs for the different forms of collaboration have led van Rijnsoever et al. (2008) to suggest specific policies for the academics of each rank. The intention of these policies would be to favor development of collaborations on the part of lower ranking academics who may lack sufficient experience, especially if they are young. On the other hand they would also stimulate maintenance of collaborations by full professors, who may not be sufficiently motivated, particularly if they are older. Theoretically, the objectives for both ranks could be achieved by more strongly flanking lower ranking academics with those of higher rank, so as to transmit them the capacity to activate and manage collaborations. However in reality, collaborations between scientists of different rank are becoming less common (Hagstrom, 1965), and they also have less impact on productivity than those between scientists of the same rank (Stvilia et al., 2011). To overcome this situation, it is necessary to improve the policies concerning collaboration, which among other things re quires development of specific indicators. In the next section we present few indicators that serve for measuring the propensity to collaborate, in the different forms, for the individual researchers that belong to each academic rank.

## 3. Methodology, dataset and indicators

*3.1 Methodology*

The analysis of rank differences in research collaboration has primarily been conducted by surveys (Melkers and Kiopa, 2010; van Rijnsoever et al., 2008; Frehill et al., 2010; Bozeman and Corley, 2004). Analysis based on bibliometric data has generally been limited to statistical validation of hypothesis, such as on the determinants of collaborations (Jeong et al., 2011; Mcdowell and Melvin, 1983; Piette and Ross, 1992; Rivellini et al., 2006; Vafeas, 2010), the effect of collaboration on promotions (Bayer and Smart, 1991), or the impact of rank on productivity of research teams (Hinnant et al., 2012; Stvilia et al., 2011; Abramo et al., 2011b). However there have apparently been no bibliometric studies to describe and analyze, in a systematic manner, the different forms of collaboration adopted by full, associate and assistant professors.

In the current article we adopt an approach to the study of co-authorship that was first proposed by the same authors (Abramo et al., 2013b), which takes the single scientist as the base unit of analysis. Specifically, the propensity to collaborate for each scientist is calculated as the ratio of their co-authored publications to the total of their publications. The propensity to collaborate in each rank will then be the average of the individual propensities of the academics belonging to that rank. This differs from the approach generally seen in the literature, which first aggregates all co-authored publications for the unit analyzed (academic rank, gender, etc.) and then divides them by the total of publications for that unit. Our approach overcomes problems connected to the distribution of scientific production, which is general very skewed[3]. The more typical aggregate measures are subject to strong distortions, with values of propensity being affected by the presence of outliers, thus not reflecting the true propensity of the large part of the academics of a given rank.

Our analysis refers to all Italian university professors in the hard sciences and some fields of the social sciences, where publications indexed by bibliometric databases represent a good proxy of overall research output (Moed, 2005). We exclude the arts and humanities, where the coverage of bibliometric databases is too limited. The instrument of analysis is the co-authorship of scientific publications over the period 2006-2010 as indexed on the WoS. For each form of collaboration, we first calculate the different propensities to collaborate for the individual scientist, then analyze differences across academic ranks, disciplines, and finally across fields within each discipline.

*3.2 Data sources and field of observation*

The dataset of Italian professors used in our analysis has been extracted from a database[4] maintained by the Ministry of Education, Universities and Research (MIUR). This database indexes the names, academic rank, discipline (UDA), field (SDS), and institutional affiliation of all academics in Italian universities.

Next, the dataset of these individuals' publications is extracted from the Italian Observatory of Public Research (ORP), a database developed and maintained by the authors and derived under license from the WoS. Beginning from the raw data of 2006-2010 Italian publications in the WoS, and applying a complex algorithm for disambiguation of the true identity of the authors and their institutional affiliations (for details see D'Angelo et al., 2011), each publication[5] is attributed to the university scientist or scientists (full, associate and assistant professors) that produced it, with a harmonic average of precision and recall of 96 (F-measure with error of 4%).

---

[3] In the Italian case 23% of academics produce 77% of overall scientific advancement (Abramo et al., 2013c).
[4] http://cercauniversita.cineca.it/php5/docenti/cerca.php, last accessed on Aug. 30, 2013.
[5] We exclude those document types that cannot be strictly considered as true research products, such as editorial material, meeting abstracts, replies, etc.

For each publication (almost 200,000 in all), the bibliometric dataset thus provides:
- the complete list of all co-authors;
- the complete list of all their addresses;
- a sub-list of only the academic authors, with their rank, SDS/UDA and university affiliations.

Our dataset permits unequivocal identification of each academic with their home university, although this operation is not possible for non-academic authors of the publications. It is also not possible to associate the academics with any organizations other than their own universities, although the literature shows (Katz and Martin, 1997) that in some cases authors indicate more than one institutional address, due to some form of multiple engagement or change in employment. This can actually lead to certain problems, such as classifying publications as being produced under international co-authorship when the presence of a foreign organization in the byline is actually due to a single academic belonging to multiple organizations (Glanzel, 2001)[6]. Further, our dataset permits unequivocal assignment of every academic to their SDS, and thus to the UDA to which they belong, while the same operation is not possible for non-academic authors of the publications. For these reasons, the analysis is conducted only for university professors.

Table 1 presents the statistics for the population of Italian academics belonging to the 11 UDAs analyzed, with their respective publications. To render the bibliometric analysis still more robust, the field of observation is limited to those SDSs (200 in all) where at least 50% of academics produce at least one publication in the 2006-2010 period.

The numbers and percentages of full, associate, and assistant professors in the composition of each UDA vary substantially. The percentage of full professors ranges from 24.6% (Medicine) to 38.1% (Economics and statistics), while that for associate professors varies from 28.3% (Economics and statistics) to 35.7% (Physics). The lowest percentage of assistant professors is registered in Physics (31.3%) and the highest is in Medicine (46.0%). These variations are primarily due to the different career policies of the various UDAs, adopted at the level of the individual SDSs and by each university administration (Lissoni et al., 2011). In particular, in Physics the number of full professors exceeds that for assistant professors, while the most numerous rank is that of associate professors. There are also substantial numbers of full professors in Agricultural and veterinary sciences, Civil engineering and Pedagogy and psychology, in all cases exceeding the number of associate professors. In Industrial and information engineering and in Economics and Statistics the full professors are actually more numerous than both the associate and the assistant professors. Thus the typical hierarchical pyramid is in fact only observed in Biology, Chemistry, Earth sciences, Mathematics, Computer sciences and Medicine.

---

[6] In Section 3.3 we describe the methodological assumptions that address this critical problem.

*Table 1: Research staff per academic rank in each UDA and relevant publications*

| UDA | Rank | Publications | Research staff† | | |
| --- | --- | --- | --- | --- | --- |
| | | | Total | Productive | Collaborative |
| Medicine (MED) | Full | 44,094 (70.0%) | 3,053 (24.6%) | 2,788 (91.3%) | 2,784 (91.2%) |
| | Associate | 31,175 (49.5%) | 3,667 (29.5%) | 3,019 (82.3%) | 3,016 (82.2%) |
| | Assistant | 28,465 (45.2%) | 5,713 (46.0%) | 4,377 (76.6%) | 4,374 (76.6%) |
| Industrial and information engineering (IIE) | Full | 22,309 (59.8%) | 2,036 (36.1%) | 1,752 (86.1%) | 1,743 (85.6%) |
| | Associate | 17,721 (47.5%) | 1,722 (30.5%) | 1,459 (84.7%) | 1,453 (84.4%) |
| | Assistant | 14,009 (37.6%) | 1,886 (33.4%) | 1,635 (86.7%) | 1,626 (86.2%) |
| Biology (BIO) | Full | 19,628 (62.8%) | 1,720 (29.4%) | 1,595 (92.7%) | 1,593 (92.6%) |
| | Associate | 13,195 (42.2%) | 1,723 (29.4%) | 1,491 (86.5%) | 1,489 (86.4%) |
| | Assistant | 13,957 (44.6%) | 2,412 (41.2%) | 2,158 (89.5%) | 2,155 (89.3%) |
| Chemistry (CHE) | Full | 17,552 (68.3%) | 1,128 (31.2%) | 1,082 (95.9%) | 1,081 (95.8%) |
| | Associate | 13,058 (50.8%) | 1,210 (33.5%) | 1,091 (90.2%) | 1,088 (89.9%) |
| | Assistant | 11,807 (46.0%) | 1,272 (35.2%) | 1,211 (95.2%) | 1,210 (95.1%) |
| Physics (PHY) | Full | 14,044 (59.3%) | 946 (32.9%) | 885 (93.6%) | 878 (92.8%) |
| | Associate | 11,558 (48.8%) | 1,027 (35.7%) | 903 (87.9%) | 891 (86.8%) |
| | Assistant | 9,396 (39.6%) | 900 (31.3%) | 814 (90.4%) | 806 (89.6%) |
| Mathematics and Computer Sciences (MAT) | Full | 8,373 (51.9%) | 1,144 (31.7%) | 980 (85.7%) | 950 (83.0%) |
| | Associate | 6,296 (39.0%) | 1,187 (32.9%) | 894 (75.3%) | 867 (73.0%) |
| | Assistant | 5,631 (34.9%) | 1,276 (35.4%) | 1,031 (80.8%) | 992 (77.7%) |
| Agricultural and veterinary sciences (AVS) | Full | 7,339 (62.4%) | 1,057 (33.2%) | 909 (86.0%) | 906 (85.7%) |
| | Associate | 5,832 (49.6%) | 908 (28.5%) | 772 (85.0%) | 770 (84.8%) |
| | Assistant | 5,976 (50.8%) | 1,218 (38.3%) | 1,039 (85.3%) | 1,038 (85.2%) |
| Pedagogy and Psychology (PPS) | Full | 2,072 (61.9%) | 321 (30.4%) | 244 (76.0%) | 242 (75.4%) |
| | Associate | 1,186 (35.5%) | 316 (30.0%) | 216 (68.4%) | 212 (67.1%) |
| | Assistant | 1,017 (30.4%) | 418 (39.6%) | 255 (61.0%) | 251 (60.0%) |
| Earth Sciences (EAR) | Full | 2,694 (51.0%) | 449 (31.6%) | 386 (86.0%) | 383 (85.3%) |
| | Associate | 2,424 (45.9%) | 487 (34.2%) | 384 (78.9%) | 383 (78.6%) |
| | Assistant | 2,019 (38.2%) | 487 (34.2%) | 411 (84.4%) | 407 (83.6%) |
| Economics and Statistics (ECS) | Full | 1,931 (54.0%) | 742 (38.1%) | 484 (65.2%) | 433 (58.4%) |
| | Associate | 1,242 (34.7%) | 552 (28.3%) | 353 (63.9%) | 328 (59.4%) |
| | Assistant | 1,020 (28.5%) | 655 (33.6%) | 363 (55.4%) | 339 (51.8%) |
| Civil engineering (CEN) | Full | 3,192 (41.4%) | 570 (32.6%) | 437 (76.7%) | 435 (76.3%) |
| | Associate | 2,221 (37.3%) | 555 (31.8%) | 374 (67.4%) | 367 (66.1%) |
| | Assistant | 2,001 (59.8%) | 622 (35.6%) | 419 (67.4%) | 407 (65.4%) |
| Total | Full | 133,570* (67.6%) | 13,166 (30.4%) | 11,542 (87.7%) | 11,428 (86.8%) |
| | Associate | 99,325* (50.3%) | 13,354 (30.8%) | 10,956 (82.0%) | 10,864 (81.4%) |
| | Assistant | 89,303* (45.2%) | 16,859 (38.9%) | 13,713 (81.3%) | 13,605 (80.7%) |

† The figures refer to the research staff working in those SDSs (200 in all) where at least 50% of academics produce at least one publication in the 2006-2010 period.
* Totals are less than the sum of the column data due to double counts of publications co-authored by full/associate/assistant professors where the research subject pertains to more than one UDA.

However the hierarchical pyramid is more closely respected in terms of the contribution to publications provided by the three academic ranks (Table 1, column 3, last 3 lines). Of the 197,460 publications in the dataset, 67.6% feature authorship by at least one full professor, compared to the 50.3% and 45.2% shares with authorship including associate and assistant professors. Descending to the UDA level, we note that the percentage of publications with associate professor authorship is exceeded by that for assistant professors in only three disciplines (Civil engineering, Agricultural and veterinary sciences, Biology). Finally, Civil engineering is the only case where the percentage of publications with assistant professor authorship exceeds that for publications with at least one full professor author. Beyond these particular cases, we thus observe a positive relationship between the rank and number of publications authored, which could be further examined in an attempt to detect the real contribution of the co-authors in the byline, including the potential occurrence of gift authorship phenomena.

The percentage of "productive" academics (producing at least one publication indexed under the WoS in the period 2006-2010) differs among full, associate and assistant professors, and at the general level is positively related to rank: in the period under examination, roughly 88% of full professors, 82% of associate professors and just over 81% of assistant professors produced at least one publication. At the level of the individual UDAs, we observe that the percentage of productive associate professors is lesser than for assistant professors, except in Economics and statistics, Medicine and Pedagogy and psychology. This result could be due to the career policies initiated in recent years, which on the one hand have favored recruitment of a high number of assistant professors, while also inserting publication in indexed journals as an essential criteria for entry into academic ranks and for subsequent progression.

Analyzing the percentage of "collaborative" academics (at least one publication in co-authorship with other scientists in the same period), both at the general level and for the individual UDAs, the conclusions are substantially the same as those noted for "productive" academics: in the large part of UDAs, the percentage of associate professors that publish exclusively alone is greater than that registered for assistant professors.

*3.3 Indicators and methods*

Beginning from the individual academics of known rank and SDS, we will compare the average propensity to collaborate in the different fields for each of four forms: general propensity, intramural propensity, and extramural with researchers from domestic and from foreign organizations. The first form, the propensity to collaborate in general, represents a superset of the others.

We construct an "author-publication" matrix of dimensions $m$ x $n$, with:
- $m = 36,211$, i.e. total number of productive academics,
- $n = 197,460$, i.e. total number of their publications.

We then associate each academic with his or her publications ($p$) over the period. Since for each publication we know the number of authors and the numbers of domestic and foreign institutions, for each scientist we can calculate the number of publications resulting from collaborations ($cp$), the number of publications resulting from collaborations with other academics belonging to the same university (intramural - $cip$), the number of publications from collaborations with scientists belonging to other domestic organizations (extramural domestic - $cedp$), and the number of publications with scientists belonging to foreign organizations (extramural international - $cefp$)[7]. From these values we can construct the indicators for the relative individual propensities to collaborate, from which we can then

---

[7] Single-authored papers with more than one affiliation are not considered as collaborations. A publication with more than two authors could present different forms of collaboration, for example intramural and extramural domestic. In this case it is counted in calculating propensity for each form of collaboration observed.

also obtain the average propensities per rank in each discipline (UDA):
- Propensity to collaborate $C = \frac{cp}{p}$
- Propensity to collaborate intramurally $CI = \frac{cip}{p}$
- Propensity to collaborate extramurally at the domestic level $CED = \frac{cedp}{p}$
- Propensity to collaborate extramurally at the international level $CEF = \frac{cefp}{p}$

Each of the four indicators varies between zero (if, in the observed period, the scientist under observation did not produce any publications resulting from the form of collaboration analyzed); and 1 (if the scientist produced all his or her publications through that form of collaboration).

## 4. Rank differences in the propensity to collaborate in different forms, in the various disciplines

The calculation of *C*, *CI*, *CED* and *CEF* permits the comparison of the values of propensity registered for full, associate, and assistant professors, relative to different forms of co-authorship and in the different UDAs. Full, associate, and assistant professors belonging to the various UDAs show different propensities to collaborate. To analyze these differences, we present a table for each form of collaboration, showing per rank and UDA: i) the average propensity to collaborate; ii) the percentage of academics with nil propensity; iii) the percentage of academics with maximum (100%) propensity. The last column of the tables show the results of the Mann-Whitney U test[8] (Mann and Whitney, 1947), which is applied to verify the significance of the observed rank differences in collaboration. The initial analysis of the differences is conducted using the Wilcox.test function (R Development Core Team, 2012); the sign + (-) in each cell highlights, for each UDA, if the academics, whose rank is indicated in the second column, have on average a higher (lower) propensity than the academics, whose rank is indicated in the penultimate column. The findings permit clustering of UDAs on the basis of the differences in propensity to collaborate of their full, associate, and assistant professors. Table 2 shows the values of propensity to collaborate, *C*. These generally appear extremely high for full, associate, and assistant professors, in line with the results obtained by Abramo et al. (2013b).

In general, the propensity for assistant professors to collaborate is higher than that for both full and associate professors. This result, further confirmed by the Mann-Whitney U test[9], is in contrast with various results presented in the literature (Vafeas, 2010; Mcdowell and Melvin, 1983; Rivellini et al., 2006), which show a lesser tendency for collaboration on the part of assistant professors.

---

[8] Although our dataset includes the entire population of Italian academics and is not a sample, we still apply the significance test for potential purposes of extending the results to other contexts and periods.

[9] The Mann-Whitney U test compares two samples, verifying the significance of the difference between the medians. For this reason there can be cases where the test shows a positive (or negative) difference between two samples even where the first sample has an average that is lower (higher) than the second (see the case of the comparison between full and associate professors in Pedagogy and psychology, Table 2).

*Table 2: Propensity to collaborate in general, C, per academic rank in each discipline (percentage values)*

| UDA* | Rank | Mean C | C = 0% | C = 100% | | U Mann-Whitney |
|---|---|---|---|---|---|---|
| CHE | Full | 99.4 | 0.1 | 94.2 | vs Associate | + |
| | Associate | 98.7 | 0.3 | 93.4 | vs Assistant | -*** |
| | Assistant | 99.4 | 0.1 | 96.7 | vs Full | +** |
| MED | Full | 99.2 | 0.1 | 91.6 | vs Associate | -*** |
| | Associate | 99.4 | 0.1 | 94.1 | vs Assistant | -*** |
| | Assistant | 99.6 | 0.1 | 97.2 | vs Full | +*** |
| AVS | Full | 99.1 | 0.3 | 95.3 | vs Associate | + |
| | Associate | 98.9 | 0.3 | 94.4 | vs Assistant | -** |
| | Assistant | 99.3 | 0.1 | 97.0 | vs Full | +* |
| BIO | Full | 99.0 | 0.1 | 92.5 | vs Associate | - |
| | Associate | 99.0 | 0.1 | 94.0 | vs Assistant | -** |
| | Assistant | 99.2 | 0.1 | 96.1 | vs Full | +*** |
| EAR | Full | 97.9 | 0.8 | 91.2 | vs Associate | + |
| | Associate | 97.5 | 0.3 | 89.3 | vs Assistant | - |
| | Assistant | 97.5 | 1.0 | 91.2 | vs Full | - |
| IIE | Full | 97.9 | 0.5 | 88.5 | vs Associate | +*** |
| | Associate | 96.5 | 0.4 | 81.4 | vs Assistant | -*** |
| | Assistant | 96.8 | 0.6 | 86.0 | vs Full | -** |
| PPS | Full | 97.4 | 0.8 | 87.7 | vs Associate | - |
| | Associate | 96.3 | 1.9 | 90.3 | vs Assistant | - |
| | Assistant | 96.3 | 1.6 | 91.4 | vs Full | + |
| PHY | Full | 97.1 | 0.8 | 79.5 | vs Associate | - |
| | Associate | 96.0 | 1.3 | 81.8 | vs Assistant | - |
| | Assistant | 96.8 | 1.0 | 83.3 | vs Full | +* |
| CEN | Full | 96.6 | 0.5 | 85.1 | vs Associate | +** |
| | Associate | 93.2 | 1.9 | 78.9 | vs Assistant | - |
| | Assistant | 93.0 | 2.9 | 80.2 | vs Full | -* |
| MAT | Full | 90.3 | 3.1 | 67.7 | vs Associate | + |
| | Associate | 88.8 | 3.0 | 68.2 | vs Assistant | - |
| | Assistant | 88.1 | 3.8 | 69.7 | vs Full | - |
| ECS | Full | 81.9 | 10.5 | 66.7 | vs Associate | - |
| | Associate | 83.6 | 7.1 | 67.4 | vs Assistant | -** |
| | Assistant | 87.1 | 6.6 | 77.1 | vs Full | +** |
| Total | Full | 97.2 | 1.0 | 87.4 | vs Associate | + |
| | Associate | 96.8 | 0.8 | 87.6 | vs Assistant | -*** |
| | Assistant | 97.5 | 0.8 | 91.4 | vs Full | +*** |

*\* See Table 1 for UDAs acronyms*

This difference could be due to the methodology of aggregate measurement, but also to the fact that Rivellini et al. (2006) analyzed collaborations exclusively limited to academics, while Vafeas (2010) and Mcdowell and Melvin (1983) focused only on the area of economics. According to these same authors, academics in this area must demonstrate their scientific capacity through publications as individual authors. In effect, from our analysis it emerges that in various UDAs (Mathematics and computer science, Civil engineering, Industrial and information engineering, Earth sciences) the percentage of scientists with a C of nil value is higher among assistant professors, yet this does not occur precisely in Economics and statistics. In this same UDA we also observe that the percentage of full and associate professors that collaborate in all publications (C = 100%) is 10 percentage points less than it is for assistant professors. The overall greater propensity for collaboration by assistant professors is confirmed in six individual UDAs, while in the remaining four (Civil engineering, Earth sciences, Industrial and information engineering, Mathematics and computer sciences), full professors achieve the maximum value of propensity to collaborate, followed by assistant professors.

The results concerning propensity to collaborate in general (*C*) do not permit identification of the possible differences in the various forms of collaboration. For this reason we deepen the analysis to the level of the different forms. The results concerning propensity for intramural collaboration, *CI*, are presented in Table 3.

*Table 3: Propensity to collaborate intramurally, CI, per academic rank in each discipline (percentage values)*

| UDA* | Rank | *Mean CI* | *CI = 0%* | *CI = 100%* | | U Mann-Whitney |
|---|---|---|---|---|---|---|
| IIE | Full | 82.0 | 3.8 | 44.6 | *vs* Associate | +*** |
|  | Associate | 79.6 | 4.6 | 39.1 | *vs* Assistant | -*** |
|  | Assistant | 84.7 | 3.4 | 56.2 | *vs* Full | +*** |
| CHE | Full | 82.0 | 1.7 | 38.3 | *vs* Associate | - |
|  | Associate | 80.5 | 3.2 | 42.1 | *vs* Assistant | -*** |
|  | Assistant | 87.6 | 2.2 | 56.7 | *vs* Full | +*** |
| AVS | Full | 80.7 | 3.7 | 50.2 | *vs* Associate | +** |
|  | Associate | 77.0 | 5.1 | 42.6 | *vs* Assistant | -*** |
|  | Assistant | 84.7 | 4.1 | 60.0 | *vs* Full | +*** |
| CEN | Full | 77.1 | 6.2 | 46.2 | *vs* Associate | +** |
|  | Associate | 70.0 | 9.4 | 41.4 | *vs* Assistant | -* |
|  | Assistant | 72.4 | 11.0 | 50.6 | *vs* Full | - |
| BIO | Full | 76.7 | 2.9 | 34.5 | *vs* Associate | - |
|  | Associate | 75.3 | 5.2 | 40.9 | *vs* Assistant | -*** |
|  | Assistant | 82.8 | 4.4 | 57.5 | *vs* Full | +*** |
| MED | Full | 76.6 | 2.9 | 29.3 | *vs* Associate | -*** |
|  | Associate | 77.0 | 4.5 | 39.2 | *vs* Assistant | -*** |
|  | Assistant | 86.7 | 3.3 | 61.1 | *vs* Full | +*** |
| PHY | Full | 67.0 | 6.9 | 29.4 | *vs* Associate | +** |
|  | Associate | 61.7 | 11.8 | 26.1 | *vs* Assistant | -*** |
|  | Assistant | 72.0 | 7.2 | 32.4 | *vs* Full | +** |
| EAR | Full | 62.0 | 10.4 | 29.3 | *vs* Associate | +* |
|  | Associate | 56.4 | 14.1 | 25.3 | *vs* Assistant | -*** |
|  | Assistant | 67.2 | 10.0 | 38.2 | *vs* Full | +** |
| PPS | Full | 60.9 | 11.1 | 30.7 | *vs* Associate | + |
|  | Associate | 55.2 | 21.8 | 31.5 | *vs* Assistant | -* |
|  | Assistant | 62.2 | 22.0 | 44.3 | *vs* Full | + |
| MAT | Full | 52.2 | 18.0 | 20.4 | *vs* Associate | - |
|  | Associate | 53.5 | 21.7 | 24.3 | *vs* Assistant | -* |
|  | Assistant | 56.6 | 21.8 | 31.0 | *vs* Full | +** |
| ECS | Full | 39.5 | 37.4 | 20.9 | *vs* Associate | - |
|  | Associate | 43.2 | 34.6 | 24.4 | *vs* Assistant | -* |
|  | Assistant | 50.8 | 35.5 | 36.6 | *vs* Full | +*** |
| Total | Full | 73.1 | 6.5 | 34.4 | *vs* Associate | + |
|  | Associate | 71.8 | 8.4 | 36.6 | *vs* Assistant | -*** |
|  | Assistant | 80.2 | 6.7 | 53.5 | *vs* Full | +*** |

* See Table 1 for UDA acronyms

For this form of collaboration, we again observe that at the general level the assistant professors show a greater propensity to collaborate. This is also confirmed for the individual UDAs, especially in Medicine, where the average propensity for intramural collaboration of assistant professors exceeds that for the other ranks by almost 10 percentage points. Civil Engineering is the only discipline in which we observe a greater propensity on the part of full professors. Excluding this case, the greater propensity for intramural collaboration on the part of assistant professors could be interpreted as a proof of the fact that academics belonging to lower ranks have less capacity to extend their collaborations beyond their home university (Hinnant et al., 2012). In fact in the Italian university system, assistant professors are often assigned to internal research groups or are entrusted to a mentor, generally of higher rank, with whom they indeed develop a large part of their collaborations, in keeping with the "faculty mentoring" system analyzed by Sands et al. (1991). The phenomena of greater networking capacity for the senior ranks also explains why full professors, who generally have a larger number of fellows (Lee and Bozeman, 2005; van Rijnsoever et al., 2008), are seen to register a slightly greater propensity for intramural collaboration than associate professors.

Table 4 presents the results for propensity to collaborate extramurally, at the domestic level. Associate professors register the greatest propensity to collaborate in this form, both overall and in six out of 11 individual UDAs, as confirmed by application of the Mann-Whitney U test. The Agricultural and veterinary sciences UDA is the only case where assistant professors register the greatest propensity for extramural domestic collaboration, although the differences with the other ranks are not significant. In the remaining four UDAs (Physics, Chemistry, Civil engineering, Mathematics and Computer Sciences), the full professors prevail. In general, the differences between ranks within each UDA are negligible, except in Pedagogy and psychology. This would explain why many of the Mann-Whitney U test results are observed to be non significant. At present we are unable to compare these results to those of other nations, due to the absence of published studies on the different extramural domestic collaboration patterns of the three ranks of professors.

The literature does offer several preceding studies on extramural collaboration at the international level (Frehill et al., 2010; Melkers and Kiopa, 2010), permitting comparison with the results obtained from the current study (Table 5). In keeping with the previous studies, we observe that full professors register the maximum propensity for international extramural collaboration. This result is confirmed both at the general level and for the individual UDAs, with the exceptions of Industrial and information engineering and Agricultural and veterinary sciences, where associate professors prevail (at least according to the Mann-Whitney U test, although the result is not actually significant). In general, the percentage of assistant professors that never collaborated at the international level is notably greater than that for full professors, while there is minimal difference between the percentages of academics that have published only in collaboration with international institutions. There are only three UDAs where there is little difference in terms of collaboration at the international level, between the percentages of assistant and full professors that have never collaborated: Civil Engineering, Agricultural and veterinary sciences and Physics, the observations for this latter discipline thus confirming the pattern outlined in the studies by Ackers (2004; 2005).

*Table 4: Propensity to collaborate extramurally at the domestic level, CED, per academic rank in each discipline (percentage values)*

| UDA* | Rank | Mean CED | CED = 0% | CED = 100% | | U Mann-Whitney |
|---|---|---|---|---|---|---|
| PHY | Full | 74.8 | 3.7 | 24.7 | vs Associate | + |
| | Associate | 72.5 | 6.3 | 26.4 | vs Assistant | +* |
| | Assistant | 70.2 | 6.8 | 23.3 | vs Full | -** |
| MED | Full | 63.8 | 4.0 | 13.6 | vs Associate | - |
| | Associate | 63.0 | 7.2 | 19.9 | vs Assistant | + |
| | Assistant | 61.0 | 11.5 | 25.6 | vs Full | - |
| EAR | Full | 57.9 | 12.2 | 21.0 | vs Associate | - |
| | Associate | 60.3 | 10.9 | 22.7 | vs Assistant | + |
| | Assistant | 57.7 | 16.1 | 26.0 | vs Full | + |
| BIO | Full | 57.5 | 6.9 | 12.2 | vs Associate | - |
| | Associate | 57.9 | 9.7 | 18.3 | vs Assistant | + |
| | Assistant | 56.9 | 12.0 | 20.8 | vs Full | + |
| CHE | Full | 50.7 | 6.7 | 7.8 | vs Associate | + |
| | Associate | 49.7 | 9.4 | 11.0 | vs Assistant | + |
| | Assistant | 49.0 | 8.6 | 9.2 | vs Full | - |
| AVS | Full | 46.8 | 16.6 | 13.0 | vs Associate | - |
| | Associate | 47.0 | 16.2 | 12.6 | vs Assistant | - |
| | Assistant | 47.3 | 18.7 | 15.6 | vs Full | + |
| PPS | Full | 42.7 | 25.8 | 11.9 | vs Associate | -** |
| | Associate | 53.2 | 23.6 | 25.5 | vs Assistant | + |
| | Assistant | 50.1 | 29.0 | 29.4 | vs Full | +* |
| ECS | Full | 37.1 | 38.4 | 15.9 | vs Associate | - |
| | Associate | 38.4 | 35.1 | 19.0 | vs Assistant | + |
| | Assistant | 38.9 | 43.3 | 23.4 | vs Full | + |
| MAT | Full | 33.9 | 28.5 | 9.0 | vs Associate | + |
| | Associate | 34.8 | 33.3 | 11.5 | vs Assistant | +* |
| | Assistant | 32.3 | 37.8 | 11.6 | vs Full | -* |
| CEN | Full | 25.1 | 41.4 | 7.1 | vs Associate | + |
| | Associate | 25.6 | 43.9 | 7.8 | vs Assistant | + |
| | Assistant | 27.3 | 47.7 | 9.8 | vs Full | - |
| IIE | Full | 25.1 | 31.3 | 5.1 | vs Associate | - |
| | Associate | 26.5 | 29.1 | 5.7 | vs Assistant | +*** |
| | Assistant | 23.0 | 38.4 | 5.1 | vs Full | -*** |
| Total | Full | 49.5 | 15.4 | 12.1 | vs Associate | -*** |
| | Associate | 51.1 | 16.0 | 16.0 | vs Assistant | +* |
| | Assistant | 50.2 | 19.2 | 18.5 | vs Full | +* |

*\* See Table 1 for UDA acronyms*

The differences in the values of *C*, *CI*, *CED* and *CEF* between full, associate and assistant professors, shown by the Mann-Whitney U test, permit identification of some differences and similarities between UDAs.

In Biology, Economics and statistics, Medicine, and Pedagogy and psychology, for collaboration at the general and intramural, it is assistant professors that have a greater propensity to collaborate. In these same disciplines, concerning extramural forms of collaboration, the associate professors prevail at the domestic level and the full professors for the international level.

Similarly in Physics and Chemistry, for collaboration at both the general and intramural level, it is again the assistant professors that have the greatest propensity, while at the extramural level it is consistently the full professors that prevail.

The other UDAs register different patterns, which cannot be readily clustered. It is interesting to note that in Civil Engineering, at the general level and for all the sub-forms of collaboration, the maximum propensity to collaborate is registered by the full professors.

The differences between full, associate and assistant professors in their forms of collaboration thus vary according to the UDAs that to which they belong. This could be due to certain factors characteristic of each UDA, such as the career policies adopted, but also due to division of research group roles among the different ranks and to the level of internationalization of the research projects.

*Table 5: Propensity to collaborate extramurally at international level, CEF, per academic rank in each discipline (percentage values)*

| UDA* | Rank | Mean CEF | CEF = 0% | CEF = 100% | | U Mann-Whitney |
|---|---|---|---|---|---|---|
| PHY | Full | 54.3 | 9.4 | 11.6 | vs Associate | +* |
| | Associate | 50.8 | 14.4 | 8.9 | vs Assistant | + |
| | Assistant | 49.9 | 12.8 | 9.5 | vs Full | -** |
| EAR | Full | 35.8 | 27.7 | 9.1 | vs Associate | +* |
| | Associate | 32.2 | 34.9 | 6.8 | vs Assistant | + |
| | Assistant | 30.7 | 39.9 | 7.8 | vs Full | -** |
| PPS | Full | 33.5 | 34.0 | 8.6 | vs Associate | + |
| | Associate | 30.0 | 39.4 | 10.2 | vs Assistant | + |
| | Assistant | 31.7 | 46.7 | 16.1 | vs Full | - |
| MAT | Full | 30.9 | 32.7 | 6.4 | vs Associate | +** |
| | Associate | 27.5 | 37.7 | 7.2 | vs Assistant | +*** |
| | Assistant | 22.8 | 47.8 | 6.0 | vs Full | -*** |
| ECS | Full | 30.1 | 48.8 | 11.8 | vs Associate | + |
| | Associate | 26.6 | 51.8 | 10.8 | vs Assistant | + |
| | Assistant | 25.1 | 57.6 | 13.8 | vs Full | -** |
| BIO | Full | 28.7 | 21.1 | 2.8 | vs Associate | +*** |
| | Associate | 26.3 | 29.4 | 2.7 | vs Assistant | + |
| | Assistant | 26.5 | 34.3 | 4.8 | vs Full | -*** |
| CHE | Full | 26.1 | 22.1 | 1.2 | vs Associate | +* |
| | Associate | 25.1 | 27.6 | 1.6 | vs Assistant | + |
| | Assistant | 24.5 | 29.3 | 1.1 | vs Full | -** |
| MED | Full | 20.2 | 29.5 | 1.8 | vs Associate | +*** |
| | Associate | 18.6 | 40.4 | 2.7 | vs Assistant | +*** |
| | Assistant | 17.5 | 50.3 | 3.8 | vs Full | -*** |
| AVS | Full | 19.4 | 43.5 | 2.4 | vs Associate | - |
| | Associate | 20.3 | 41.6 | 3.0 | vs Assistant | + |
| | Assistant | 20.6 | 45.8 | 4.0 | vs Full | + |
| CEN | Full | 14.4 | 57.4 | 2.1 | vs Associate | + |
| | Associate | 15.2 | 60.4 | 3.2 | vs Assistant | - |
| | Assistant | 16.2 | 60.4 | 3.3 | vs Full | - |
| IIE | Full | 13.5 | 47.9 | 1.3 | vs Associate | - |
| | Associate | 13.5 | 47.1 | 1.2 | vs Assistant | +** |
| | Assistant | 13.1 | 54.9 | 2.0 | vs Full | -** |
| Total | Full | 25.4 | 32.2 | 3.8 | vs Associate | +*** |
| | Associate | 24.0 | 37.1 | 3.9 | vs Assistant | +*** |
| | Assistant | 22.4 | 43.8 | 4.6 | vs Full | -*** |

* See Table 1 for UDAs acronyms

# 5. Conclusions

For academics, it is becoming ever more important to establish research collaborations with colleagues in their own and other universities and with other domestic and international institutions. Collaborations permit participation in broader research projects, access to funding, and not least, improvement in personal competencies, with positive effects on the quantity and quality of publications. Such results also give rise to evident advantages for the universities and for national research systems, and in fact these increasingly insert the capacity to activate and manage effective collaborations among the criteria for promotion to higher ranks.

In spite of the fact that such policies have been active for a number of years, the studies concerning the relationship between rank and scientific collaboration have been focused on the differences between untenured and tenured researchers and students. In rare cases there has been an analysis of the situation of tenured researchers belonging to different academic ranks, but these have dealt only with their collaboration at the general level or through one or two selected forms. For this reason, the literature does not permit development of a systematic framework of the relationships between academic rank and forms of collaboration, in spite of the fact that the scientific profile and the functions associated with the different ranks, by regulation and academic tradition, would lead us to expect advantages and specializations of full, associate and assistant professors in specific forms of collaboration.

The present study assists in resolving the gap in the literature. Adopting an approach beginning from the single scientist as the base analytical unit, we are able to measure the propensity of the individual research to activate each form of collaboration, in an accurate manner. This permits obtainment of a realistic depiction of the behavior of the researchers belonging to the context analyzed, and the conduct of comparisons between full, associate and assistant professors that are not distorted by the presence of outliers, where individual scientists alone develop a high number of collaborations and thus heavily weigh on the value of indexes based on aggregate measures.

The application of our methodology to the scientific production of Italian academics has shown that, at the general level (not considering individual disciplines), the differences between the three ranks of professors are notably marked concerning intramural collaboration and extramural collaboration at the international level. For intramural collaboration, the analysis demonstrates a clearly greater propensity on the part of assistant professors, while for international collaboration, full and associate professors both prevail, but in a less defined manner. This illustrates how lower ranking academics tend to construct their collaborations above all with colleagues at their home university, since at this stage they are less able to activate external collaborations with the same intensity as higher ranking colleagues. This could be due to the lack of experience and visibility of the assistant professors, particularly for the youngest, but also to the division of roles within research groups, which generally leads the higher ranking academics to be involved in international collaborations. This situation should be provided for in national research assessment exercises, which sometimes include indications for intensity of international collaboration, to the detriment of assistant professors and the institutions where their presence is concentrated.

On the other hand, it is only in a few disciplines, such as Physics, that the analysis registers very high values of propensity to collaborate internationally for the assistant professors, although still lesser than for the higher ranks. The influence of the discipline on the relations between rank and forms of collaboration is further shown through the comparison of the propensities to collaborate in the different forms, for the full, associate and assistant professors across UDAs. In fact, we observe a fairly notable level of heterogeneity between the disciplines that can be explained in light of the different career policies adopted, in the division of research group roles between academic of different ranks, and in the level of internationalization of research projects.

To verify the impact of discipline on the relations between academic rank and forms of collaboration, it would useful to conduct a deeper investigation of the determinants of the different forms of collaboration, while taking into consideration other variables such as gender and the geographic position of the home university, the influence of which has already been noted in the literature. Another useful inquiry would involve analyzing the composition of the collaborations, particularly highlighting the rank of the different co-authors and calculating the propensity of the academics to collaborate with colleagues of the same rank or of different rank. This would then permit evaluation of whether collaborations with full professors lead to advantages for the lower ranked assistants, particularly in terms of productivity, as has been suggested in the literature. The authors will examine these more detailed questions in further research.